\documentstyle[12pt,epsfig]{article}

\begin{document}  
\vspace*{-2cm}  
\renewcommand{\thefootnote}{\fnsymbol{footnote}}  
\begin{flushright}  
hep-ph/9810432\\
DTP/98/66\\  
October 1998\\  
\end{flushright}  
\vskip 65pt  
\begin{center}  
{\Large \bf Heavy Quark Production at a $\gamma\gamma$ Collider:
the Effect of Large Logarithmic Perturbative Corrections}\\
\vspace{1.2cm} 
{\bf  
Michael Melles${}^1$\footnote{Michael.Melles@durham.ac.uk} and  
W.~James~Stirling${}^{1,2}$\footnote{W.J.Stirling@durham.ac.uk}  
}\\  
\vspace{10pt}  
{\sf 1) Department of Physics, University of Durham,  
Durham DH1 3LE, U.K.\\  
  
2) Department of Mathematical Sciences, University of Durham,  
Durham DH1 3LE, U.K.}  
  
\vspace{70pt}  
\begin{abstract}
We present quantitative 
results on the  cross section for $\gamma + \gamma
\left( J_z=0 \right) \longrightarrow b \bar{b}\longrightarrow$ two $b$--jets
 based on the recently
achieved all orders resummation of large soft (Sudakov) and hard (non-Sudakov)
double logarithms. The next--to--leading order QCD perturbative
 corrections are included exactly.
We find that one needs to include at least four loops for the novel
hard leading logarithms, on  the cross section level, in order to be safe
from large opposite sign cancellations that plagued earlier phenomenological
studies.
We find that the background to intermediate mass
Higgs boson production at a future photon linear collider  (PLC)
 is thus reasonably well under control and should
allow the direct determination of the partial Higgs width $\Gamma \left( H 
\longrightarrow \gamma + \gamma \right)$. Assuming high efficiency $b$-tagging,
the total Higgs width measurement at a PLC is thus a realistic goal.
\end{abstract}
\end{center}  
\vskip12pt

\setcounter{footnote}{0}  
\renewcommand{\thefootnote}{\arabic{footnote}}  
  
\vfill  
\clearpage  
\setcounter{page}{1}  
\pagestyle{plain} 
 
\section{Introduction} 

The measurement of the total Higgs width is one of the most important goals
of a future $\gamma \gamma$ collider. Besides providing various other physics
opportunities, such as the determination of the CP eigenvalues of a Higgs
boson, the photon linear collider (PLC) offers so far the only possibility
of a direct measurement of the partial width $\Gamma \left( H \longrightarrow
\gamma + \gamma \right)$ for an intermediate mass Higgs particle \cite{j1}.
With the knowledge of the respective branching ratio BR$\left( H 
\longrightarrow \gamma + \gamma \right)$ from a Higgs production process at
some earlier collider experiment \cite{egn,ik}, the total Higgs width can be
reconstructed. This in turn permits a model independent determination of
various other partial widths.

Using Compton backscattering \cite{g1,t1} of initially polarized electrons and
positrons at a linear collider, the dominant background for a Higgs boson below
the $W^\pm$ threshold is from $\gamma + \gamma  \left( J_z=0 \right)
\longrightarrow b + \overline{b}$. While this background is suppressed by
$\frac{m_b^2}{s}$ at the Born level (unlike the $J_z = \pm 2$ background which
has no such mass suppression), 
higher--order perturbative QCD corrections remove this suppression \cite{bkso}. 
In addition,
very large virtual (non-Sudakov) double logarithms (i.e.
$\log^2(s/m_b^2)$) are present at each order in perturbation theory, and 
at one loop these
can lead to a {\it negative} cross section in the central production region 
where the Higgs signal is expected \cite{jt1}. 
In Ref.~\cite{fkm} it was shown that positivity of the cross section is restored
by the leading two--loop non--Sudakov logarithmic 
corrections and recently, in Ref.~\cite{ms},
these corrections were resummed to all orders. 

The problem with previous \cite{bkso,jt1} phenomenological studies 
was that the large higher--order leading logarithmic corrections were 
not known and therefore not included. 
It is the purpose of this study to quantify the effect of these higher--order
logarithms on the cross section. We include all the results presently 
available ---  the exact one--loop correction  \cite{jt1} and the leading
double logarithms to all orders --- to make more reliable predictions for
the Higgs background.

We do not attempt here a full Monte Carlo study of the signal 
and background taking all hadronization and detector effects
into account. Instead, we note that the dominant configuration
of the Higgs signal process is a pair of back--to--back jets, 
produced centrally,  each containing a $b$ quark that can in principle
be tagged using a vertex detector. We therefore calculate the background cross 
section for the same topology. Note that this means  excluding 
multi($\geq3$)--jet topologies and also two-jet configurations arising
from Compton-like $2\to 3$ scatterings, as studied in detail
 in Ref.~\cite{bkso}, where only one of the two jets contains a $b$ quark.
 
For simplicity, we normalize all our cross sections to that of 
the leading order
$\gamma + \gamma  \left( J_z=0 \right)
\longrightarrow b + \overline{b}$ process. The Born amplitude for this is
given by 
\begin{equation}
{\cal M}_{\rm Born}(\lambda_\gamma,\lambda_q)
= \frac{8 \pi \alpha Q_q^2}{(1-\beta^2 \cos^2 \theta)} \frac{2 m_q}{
\sqrt{s}} \left(\lambda_\gamma+ \lambda_q\beta \right) \label{eq:Ba}
\end{equation}
where $\lambda_\gamma$ and $\lambda_q$ label the helicities of the photon and
quark respectively.
The  Born cross section for the $\left( J_z=0 \right)$ helicity state is then
\begin{equation}
\frac{d \sigma_{\rm Born}} {d \cos \theta}
\left( \gamma + \gamma \longrightarrow
q + \overline{q} \right) = \frac{12 \pi \alpha^2 Q_q^4}{s} \frac{
\beta \left( 1 - \beta^4 \right)}{ \left(1 - \beta^2 \cos^2 \theta \right)^2}
\label{eq:Bdcs}
\end{equation}
where $\beta=\sqrt{1-\frac{4 m_q^2}{s}}$ denotes the quark velocity.
Here $Q_q$ is the
charge of the quark with mass $m_q$, $\alpha=\frac{1}{137}$ is the fine structure
constant, $\theta$ is the quark scattering angle in the 
center--of--mass frame, 
and $\sqrt{s}$ is the overall center--of--mass scattering energy.

The paper is structured as follows.
We begin in the next section by summarizing the results
of Ref.~\cite{ms} and discussing the real gluon emission contributions.
Numerical results are given in Section~\ref{sec:nr} 
and we make concluding remarks in Section~\ref{sec:sum}.

\section{Higher--order Corrections}

The Born cross section for polarized 
$\gamma \gamma \left( J_z=0 \right)\to q \overline{q}$  collisions
given in Eq.~(\ref{eq:Ba}) receives large $\sim (\alpha_s \log^2(s/m^2))^n$
corrections at each order in perturbation theory. These `novel' 
hard\footnote{We use the description `hard' to indicate that it 
is the heavy quark mass which acts as the effective
infrared cutoff for the new double logarithms.} 
double logarithms arise from corrections which are effectively cut off by
{\it quarks}, rather than soft
gluons which generate the usual Sudakov double logarithms.
In Ref.~\cite{ms} we showed how these new double logarithmic (DL) corrections
can be resummed to all orders in perturbation theory. The
result takes the form of a confluent hypergeometric function $_2F_2$ which
possesses a $\log \left( \frac{\alpha_s}{\pi} \log^2 \frac{m^2}{s} \right)$
high energy limit. Taking into account the $\frac{m}{\sqrt{s}}$ suppression
contained in the Born amplitude (\ref{eq:Ba}), the new corrections are 
well behaved as
$s \longrightarrow \infty$. The series expansion agrees with the known one-- and
two--loop results of Refs.~\cite{jt1,fkm} as well as our 
explicit three--loop calculation \cite{ms}.

It was furthermore shown in \cite{ms} 
that at three loops all additional doubly logarithmic
corrections are described by the exponentiation of Sudakov logarithms
at each order in the expansion of the new hard form factor. This behavior
is already present at one and two loops and can thus safely be extrapolated
to all orders.

For completeness,  we list here
the resulting  virtual DL--form factor contribution to the amplitude
for the process 
$\gamma + \gamma \left( J_z=0 \right) \longrightarrow q + \overline{q}$:
\begin{eqnarray}
{\cal M}_{\rm DL} &=& {\cal M}_{\rm Born} \left\{ \exp \left( {\cal F}_{\cal A} \right) +
{\cal F} \;\; 
_2F_2 (1,1;2,\frac{3}{2}; \frac{1}{2} 
{\cal F} ) +
2 \; {\cal F} \;\; 
_2F_2 (1,1;2,\frac{3}{2}; \frac{C_A}{4 C_F} 
{\cal F} ) \right. \nonumber \\
&& \;\;\;\;\;\;\;\;\;\;\;\;\;\; + {\cal F} \;\; 
_2F_2 (1,1;2,\frac{3}{2}; \frac{1}{2} 
{\cal F} ) \; \left[  \; \exp \left( {\cal F}_{\cal A} \right) -1 \right] 
\nonumber \\
&& \left. \;\;\;\;\;\;\;\;\;\;\;\;\;\; + 2 \; {\cal F} \;\; 
_2F_2 (1,1;2,\frac{3}{2}; \frac{C_A}{4 C_F}  
{\cal F} ) \; \left[  \; 
\exp \left( {\cal F}_{\cal A} \right) -1
\right] \right\} \\
&=& {\cal M}_{\rm Born} \left\{ 1 +
{\cal F} \;\; 
_2F_2 (1,1;2,\frac{3}{2}; \frac{1}{2} 
{\cal F} ) +
2 \; {\cal F} \;\; 
_2F_2 (1,1;2,\frac{3}{2}; \frac{C_A}{4 C_F} 
{\cal F} ) \right\} 
\exp \left( {\cal F}_{\cal A} \right) \nonumber 
\label{eq:DL}
\end{eqnarray}
where 
\begin{equation}
{\cal F}_{\cal A} \equiv -C_F \frac{\alpha_s}{2 \pi} \left( \frac{1}{2} \log^2
\frac{m^2}{s} + \log \frac{m^2}{s} \log \frac{\lambda^2}{m^2} \right) \label{eq:
FA}
\end{equation}
\begin{equation}
{\cal F} \equiv -C_F \frac{\alpha_s}{4 \pi} \log^2
\frac{m^2}{s} \label{eq:F}
\end{equation}
denote the soft and hard one--loop
form factors, respectively. A fictitious gluon mass $\lambda$ is introduced
to regulate the infrared divergences in the former.  Of course
in a physical cross section, the soft
form factor ${\cal F}_{\cal A}$ cancels 
the corresponding infrared divergent contributions from the emission of real soft
gluons. For the two--jet--like contributions considered in this work, i.e.
soft gluon radiation below an energy cut of $k_g \leq \epsilon\sqrt{s}$
and arbitrarily hard gluons collinear with one of the b-quarks (specified
by a cone of half--angle $\delta$), we need to make sure that the 
 jet definition does not
restrict the exponentiation of the energy cut dependent piece of the soft
gluon matrix elements. In this case we are able to give the contribution
of the real soft DL corrections to all orders by writing
\begin{equation}
\frac{d \sigma^{DL}_{{\rm virt}+{\rm soft}}}{d \cos \theta} 
\sim \vert{\cal M}\vert^2_{\rm DL} 
\exp \left\{-2 {\cal F}_A + \Delta_c \right\}
\end{equation}
where the last term in the exponential $\Delta_c$ depends on the soft--gluon
cut  prescriptions.
Restricting the gluon energies by $E_g \leq k_c$, for example\footnote{One could also
use an invariant mass (`$y_{\rm cut}$') type cutoff.}, gives to leading
logarithmic order,
\begin{equation}
\label{eq:deltac}
\Delta_c=- \frac{ \alpha_s C_F}{\pi} \log \frac{s}{m_q^2} \log \frac{m_q^2}{4 k_c^2}
\end{equation}
At one loop, however, such DL expressions are insufficient as the 
integrated real gluon contributions also include {\it  subleading} logarithms 
with a cut dependence.
For our jet definition, we are actually able to employ exact expressions for the
infrared divergent functions as these 
have the same structure as in the QED case
\cite{yfs,m1}. We therefore use (assuming only ${m_q^2}/{s} \ll 1$):
\begin{equation}
\Delta_c= - \frac{ \alpha_s C_F}{\pi} \left( \log \frac{s}{m_q^2} \log \frac{m_q^2}
{4 k_c^2} - \log \frac{m_q^2}{4 k_c^2} + \frac{\pi^2}{3} \right) 
\end{equation}
({\it cf.} Eq.~(\ref{eq:deltac})), which leads to the physical cross section for
virtual and soft ($E_g < k_c$) real gluon emissions:
\begin{eqnarray}
\sigma^{DL}_{{\rm virt}+{\rm soft}} &=& \sigma_{\rm Born}
\left\{ 1 +
{\cal F} \;\; 
_2F_2 (1,1;2,\frac{3}{2}; \frac{1}{2} 
{\cal F} ) +
2 \; {\cal F} \;\; 
_2F_2 (1,1;2,\frac{3}{2}; \frac{C_A}{4 C_F} 
{\cal F} ) \right\}^2 \nonumber \\
&& \exp \left( \frac{ \alpha_s C_F}{\pi} \left[ \log \frac{s}{m_q^2} \left(
\frac{1}{2} - \log \frac{s}{4 k_c^2} \right) + \log \frac{s}{4 k_c^2} -1 + 
\frac{\pi^2}{3} \right] \right) \label{eq:vps}
\end{eqnarray}
\begin{figure}[t]
\centering
\epsfig{file=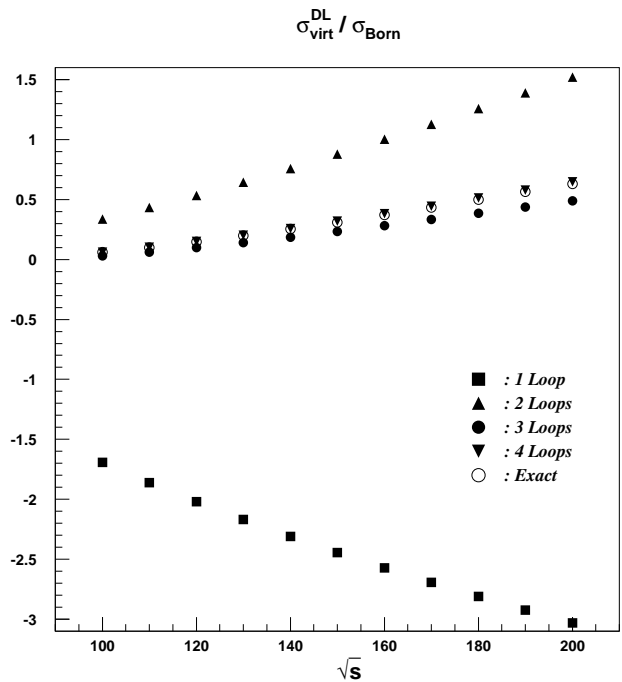,height=10cm}
\caption{The size of the virtual double logarithmic (DL) contributions relative to
the Born cross section through four loops. The `exact' result 
(open circles) is given
by the all orders resummation according to Eq.~(\ref{eq:vps}) 
and is in very good agreement with the four--loop approximation 
given in Eq.~(\ref{eq:sDL}). The huge one and two
loop contributions can be seen to lead to physically distorted results.}
\label{fig:dl}
\end{figure}
When combined with hard ($E_g>k_c$) real gluon emission appropriate to some 
particular (i.e. two--jet--like) final state topology, the logarithmic
dependence on the cutoff $k_c$ will cancel and the Sudakov form factor
will become ${\cal O}(1)$. The remaining double logarithmic corrections
then come entirely from the hard $\{\}^2$ non--Sudakov form factor.
In order to demonstrate the numerical impact of these
logarithms we list the expansion of the $\{\}^2$ piece
 in Eq.~(\ref{eq:DL}). 
Using  the coefficients in the series expansion
 for the hypergeometric function,
we obtain the expansion through four loops:
\begin{eqnarray}
\frac{\sigma^{\rm DL}_{\rm virt}}{\sigma_{\rm Born}} &\sim& 1 + 6 {\cal F}
+\frac{1}{6} \left(56 +2 \frac{C_A}{C_F} \right) {\cal F}^2
+\frac{1}{90} \left( 94 +90 \frac{C_A}{C_F} +2 \frac{C_A^2}{C_F^2} \right)
{\cal F}^3 \nonumber \\
&& +\frac{1}{2520} \left( 418 + 140 \frac{C_A}{C_F} + 238 \frac{C_A^2}{C_F^2}
+3 \frac{C_A^3}{C_F^3} \right) {\cal F}^4 + {\cal O} \left( {\cal F}^5 \right)
\label{eq:sDL}
\end{eqnarray}
where ${\cal F} = - C_F \frac{\alpha_s}{4 \pi} \log^2 \frac{m^2}{s}$
is again the one--loop hard form factor. 
Figure~\ref{fig:dl} shows the respective
contributions\footnote{Note that here the `$n$--loop contribution'
means the sum of the contributions up to and including the
$n^{\rm th}$ order contribution in Eq.~(\ref{eq:sDL}).} of the 
 terms listed in Eq.~(\ref{eq:sDL}) relative 
to the Born cross section (\ref{eq:Bdcs}). For illustration, we use parameter
values of $\alpha_s = 0.11$ and $m \equiv m_b = 4.5$~GeV, so that
${\cal F}$ varies between $-0.45$ at $\sqrt{s}=100$~GeV
and $-0.67$ at $\sqrt{s}=200$~GeV
The large cancellations between the lower order terms are clearly 
visible and only the four--loop expansion is close to the exact resummed result.
The relative size of the doubly logarithmic three-- and four--loop corrections with
respect to the full answer are depicted in Fig.~\ref{fig:relff}. While taking only
the three--loop expansion leads to a deviation of up to 50$\%$ compared to the
exact result, the four--loop DL cross section (\ref{eq:sDL}) stays within a few 
percent.
In the next section we present the results based on inclusion of the full 
one--loop radiative corrections including the full two--jet--like 
bremsstrahlung contribution.
\begin{figure}[t]
\centering
\epsfig{file=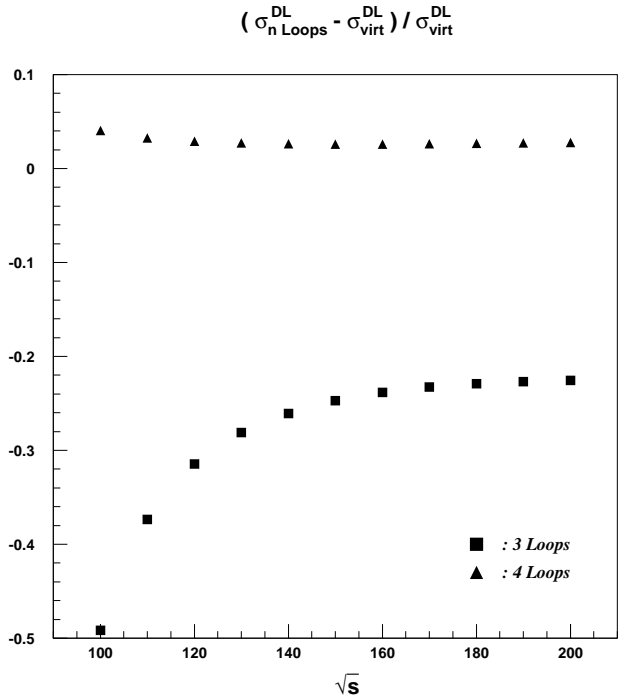,height=10cm}
\caption{The relative size of the virtual double logarithmic
contributions through three (squares) and four (triangles) loops.
With the expected experimental precision
\protect\cite{bkso} one needs to include at least four--loop corrections or
simply the resummed
values according to Eq.~(\protect\ref{eq:vps}).}
\label{fig:relff}
\end{figure}

\section{Numerical Results}
\label{sec:nr}

The results of the previous section confirm that the hard double logarithms
{\it are} numerically important when one is considering $b \overline{b}$ production
in the $\sqrt{s} = 100 - 200$~GeV energy range. The one--loop contribution
on its own tends to drive the cross section negative, and stability is
only achieved at four--loop order in the cross section. 
However before drawing definite conclusions,
it is important to assess the other contributions to the physical cross section,
in particular (i) the sub--leading logarithmic corrections to the virtual $+$
soft form factors and (ii) the contributions from hard gluon emission.
Unfortunately at present we know nothing about these beyond first order.
However the first--order result {\it is} worth studying in detail, since it allows
us to assess the dominance or otherwise of the `$6{\cal F}$' hard double
logarithm term in (\ref{eq:sDL}).

We first need to define an infrared safe two--jet cross section.\footnote{Note
that all cross sections discussed and calculated here correspond to the 
$J_Z = 0$ $\gamma\gamma$ polarization state.} As discussed
in the Introduction, our interest here is in two jet like configurations,
where each jet contains a $b$ quark. It is convenient, for purposes of 
illustration, to use a modification of 
the Sterman--Weinberg (SW) two jet definition \cite{sw}. At leading order
(i.e. $\gamma\gamma \to b \overline{b}$) all events obviously
satisfy the two--$b$--jet
requirement.  We apply an angular cut of
$\vert{\cos\theta_{b,\overline{b}}}\vert
< 0.7$ to ensure that both jets lie in the central region. This defines
our `leading order' (LO) cross section. 
At next--to--leading order (NLO) we can have virtual or real gluon emission.
For the latter, an event is defined as two--$b$--jet like if the emitted
gluon 
\begin{eqnarray*}
\mbox{{\it either}}&& \mbox{I.\quad has energy less than $\epsilon \sqrt{s}$, with  
$\epsilon \ll 1$}, \\
\mbox{{\it or}} && \mbox{II.\quad is within an angle 
$2 \delta$ of the $b$ or $\overline{b}$, again
with $\delta \ll 1$}.
\end{eqnarray*}
We will call the two regions of phase space I and II respectively.  
We further subdivide region I according to whether
the gluon energy is greater or less than the infrared cutoff\footnote{Note
that unlike $\epsilon$, $k_c$ is an unphysical parameter introduced
simply to separate soft and hard real emission.} $k_c$
 ($< \epsilon$). We combine the  region  of soft real gluon 
 phase space $E_g < k_c$ with the virtual gluon contribution to give 
 $\sigma_{\rm SV}$. The region $\tilde{{\rm I}}$ of `hard' real gluon 
 phase space $ k_c < E_g <  \epsilon\sqrt{s} $ then defines 
 $\sigma_{\tilde{\rm I}}$.
The total NLO two--$b$--jet cross section is then
\begin{equation}
\sigma_{\rm NLO} = \sigma_{\rm SV} + \sigma_{\tilde{\rm I}} + 
\sigma_{\rm II}
\end{equation}

\begin{figure}[t]
\centering
\epsfig{file=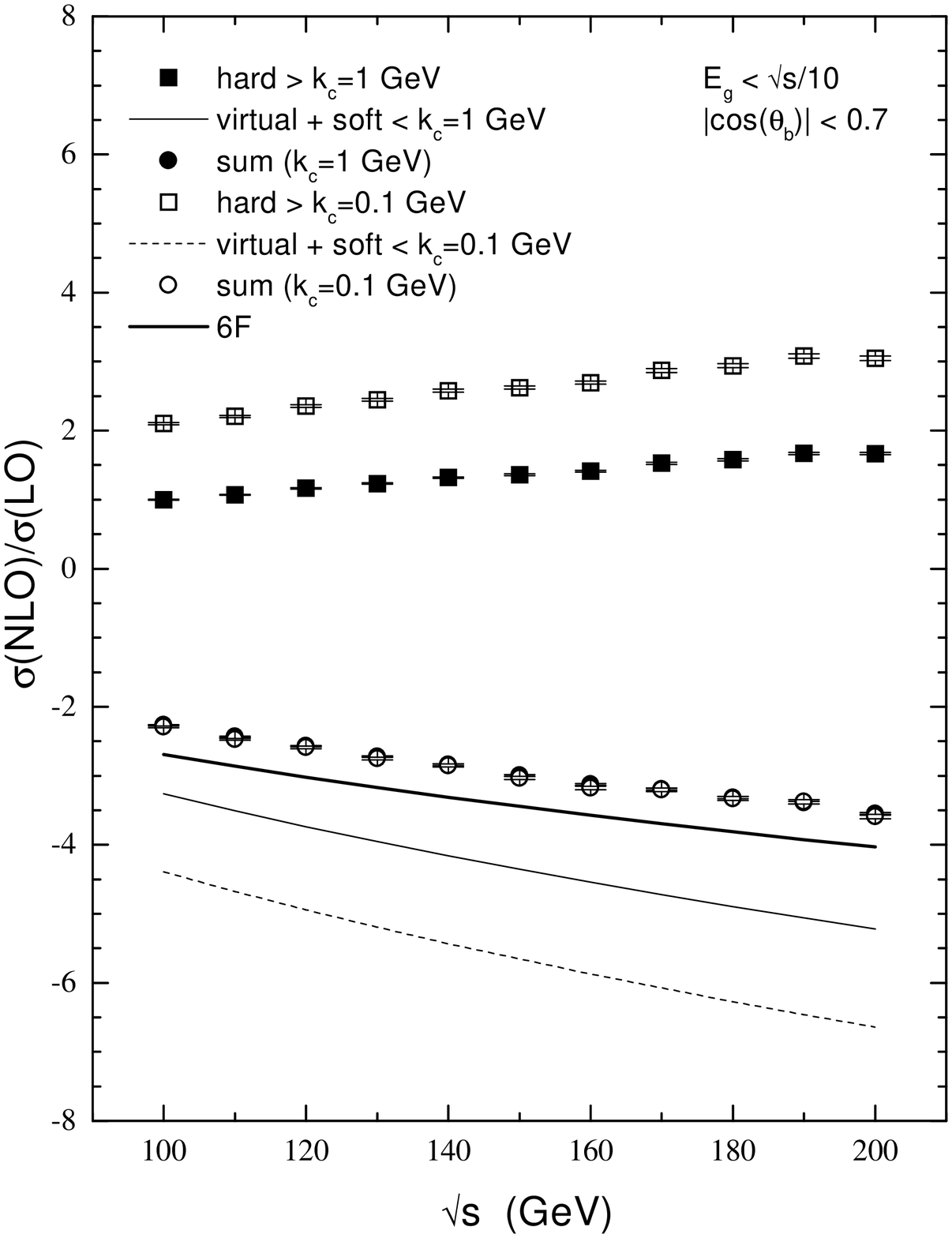,height=16cm}
\caption{Virtual and real ($E_g < k_c$, $k_c < E_g < 0.1
\protect\sqrt{s}$)
contributions to the exact next--to--leading order
two--$b$--jet cross section defined in the text. The hard double logarithm
contribution at this order ($6{\cal F}$) is also shown.}
\label{fig:insens3}
\end{figure}

The cross section  for $\sigma_{\rm SV}$ is obtained  from 
the analytic expression given in \cite{jt1},
while $\sigma_{\tilde{\rm I}}$ and  
$\sigma_{\rm II}$ are computed using the numerical program of
\cite{bkso}. A powerful consistency check on the overall calculation is that
the sum $\sigma_{\rm SV} + \sigma_{\tilde{\rm I}}$ should be independent
of the unphysical parameter $k_c$. This is 
demonstrated in Fig.~\ref{fig:insens3}, which
shows the individual contributions as a function of $\sqrt{s}$ for two choices,
$k_c = 1$~GeV and $0.1$~GeV and with $\epsilon = 0.1$. 
While the individual contributions are of course different  (the leading 
behavior as $k_c \to 0$ is $\sim \pm\log(s/m_q^2)\log(s/k_c^2)$ for
the separate SV and  ${\tilde{\rm I}}$ contributions), the sums
are indistinguishable. We can therefore conclude that for this part of the 
two--jet phase space (${\rm SV} + {\tilde{\rm I}}$), 
the next--to--leading order cross section is between $-2$ and $-3$
times the leading order cross section.

Also shown in Fig.~\ref{fig:insens3} is the  `$6{\cal F}$' hard double
logarithm contribution. Quite remarkably, this is close to the complete
result, demonstrating that the net effect of the additional 
subleading real and virtual contributions is small and positive, 
at least for this choice of parameters (in particular $\epsilon$).
In fact to a very good approximation, in this energy range,
\begin{equation}
\sigma_{\rm SV} + \sigma_{\tilde{\rm I}} = \sigma_{\rm LO}\;
\left[ 6{\cal F} + 0.3 + 0.001\sqrt{s}\; ({\rm GeV})\right]
\end{equation}
We may conclude that including the higher--order terms in the
hard form factor {\it does} give an improved prediction for the cross section,
and in particular restores positivity for this region of two--jet
phase space.

\begin{figure}[t]
\centering
\epsfig{file=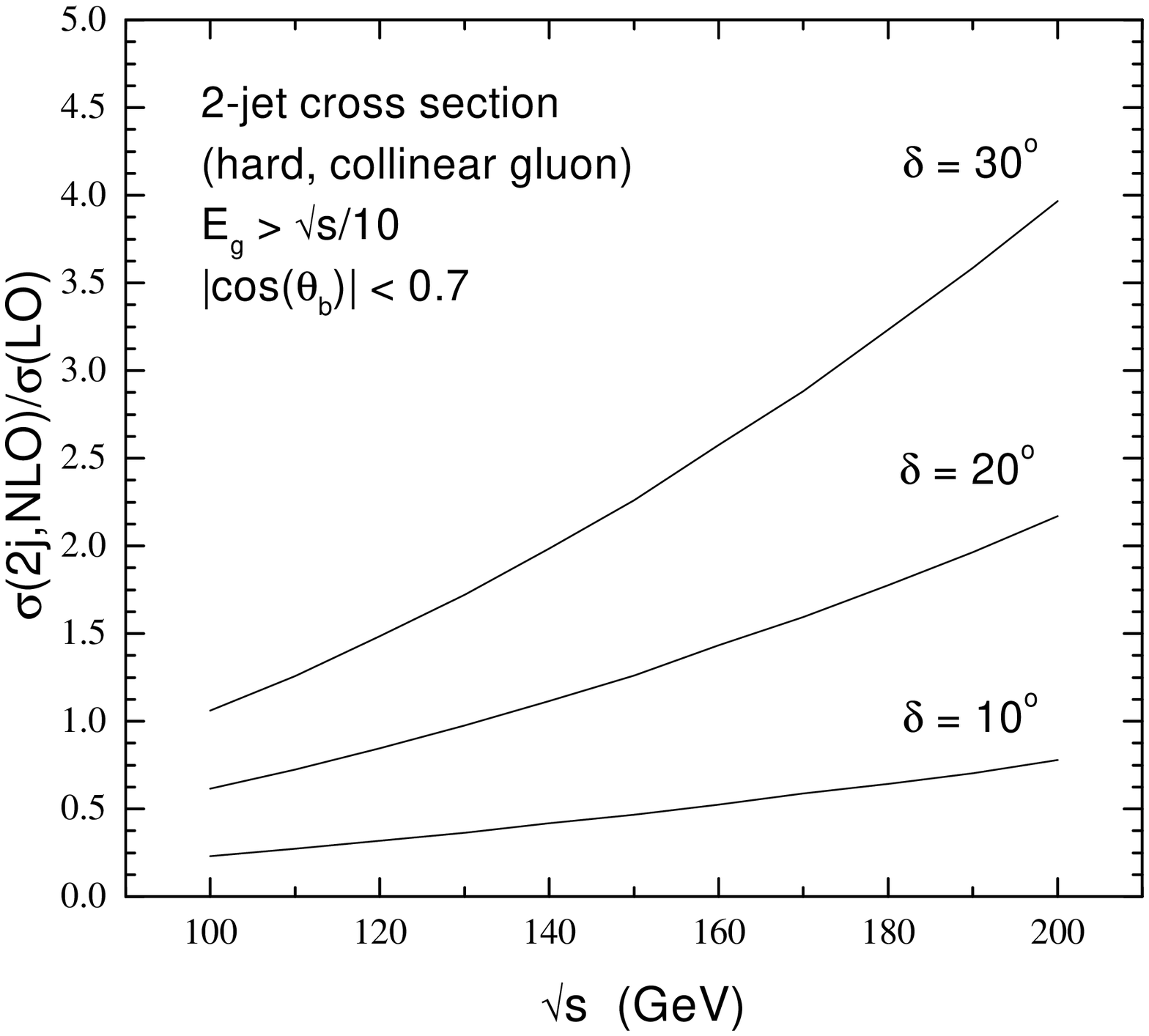,width=12cm}
\caption{Variations of the one--loop real bremsstrahlungs contribution to the cross
section. One can clearly see the removal of the $\frac{m_q^2}{s}$ suppression as
the available phase space is enlarged.}
\label{fig:insens4}
\end{figure}
The remaining part of the two--$b$--jet cross section, $\sigma_{\rm II}$,
is shown relative to $\sigma_{\rm LO}$ 
in Fig.~\ref{fig:insens4}, for three choices of the cone size $\delta$.
Consistent with the results presented in Fig.~\ref{fig:insens3}, the parameter
$\epsilon$ is again fixed at 0.1. Notice that this part of the
two--jet cross section is large and positive and approximately linear
in $\delta$.  The reason for this was first pointed out
in Ref.~\cite{bkso}: for $m_q \ll E_g \ll \sqrt{s}$ the $q\overline{q}g$
cross section behaves as 
\begin{equation}
{d\sigma \over dE_g}(\gamma \gamma \to q\overline{q}g, J_z=0) \; \sim\;
  \alpha^2 \, \alpha_s \,
{E_g^3 \over s^3} \;  \left[ \ldots \right]   
\end{equation}
i.e. with {\it no} $m_q^2/s $ suppression. This part of the 
cross section is therefore dominated by hard gluon emission, i.e. 
the $b$ quark in one of the two $b$--jets is likely to be soft.
The non--$m_q^2/s $--suppressed cross section has no collinear
singularity either, and so the dependence on the cone size $\delta$
is simply determined by phase space, i.e. $\sigma_{\rm II} \sim
O(\delta)$, for $\delta \ll 1$. 
An efficient suppression of the background to the Higgs signal
will therefore necessitate selecting narrow $b$ jets in which
the heavy quark carries a large fraction of the jet momentum.

\section{Summary and Conclusions}
\label{sec:sum}

In the previous section we have studied two types of correction
to the ($J_z=0$) $\gamma\gamma \to q \overline{q}$  cross section: the
all--orders resummed hard quark mass double logarithms, and the exact
next-to-leading order (one--loop) corrections. Motivated by the topology
of the Higgs signal, to which our contributions are a background,
we have focused on the two--$b$--jet cross section, defined here
by two parameters, an `energy outside the cone' parameter $\epsilon$
and a cone size parameter $\delta$. Furthermore we have shown that
the part of the NLO cross section corresponding to virtual and soft real
gluon emission  is dominated by the leading double hard logarithm.
This emphasizes the importance of resumming these contributions.
On the other hand, the hard collinear gluon part of the cross section
is sizable and depends quite sensitively on the jet parameters.

\begin{figure}[t]
\centering
\epsfig{file=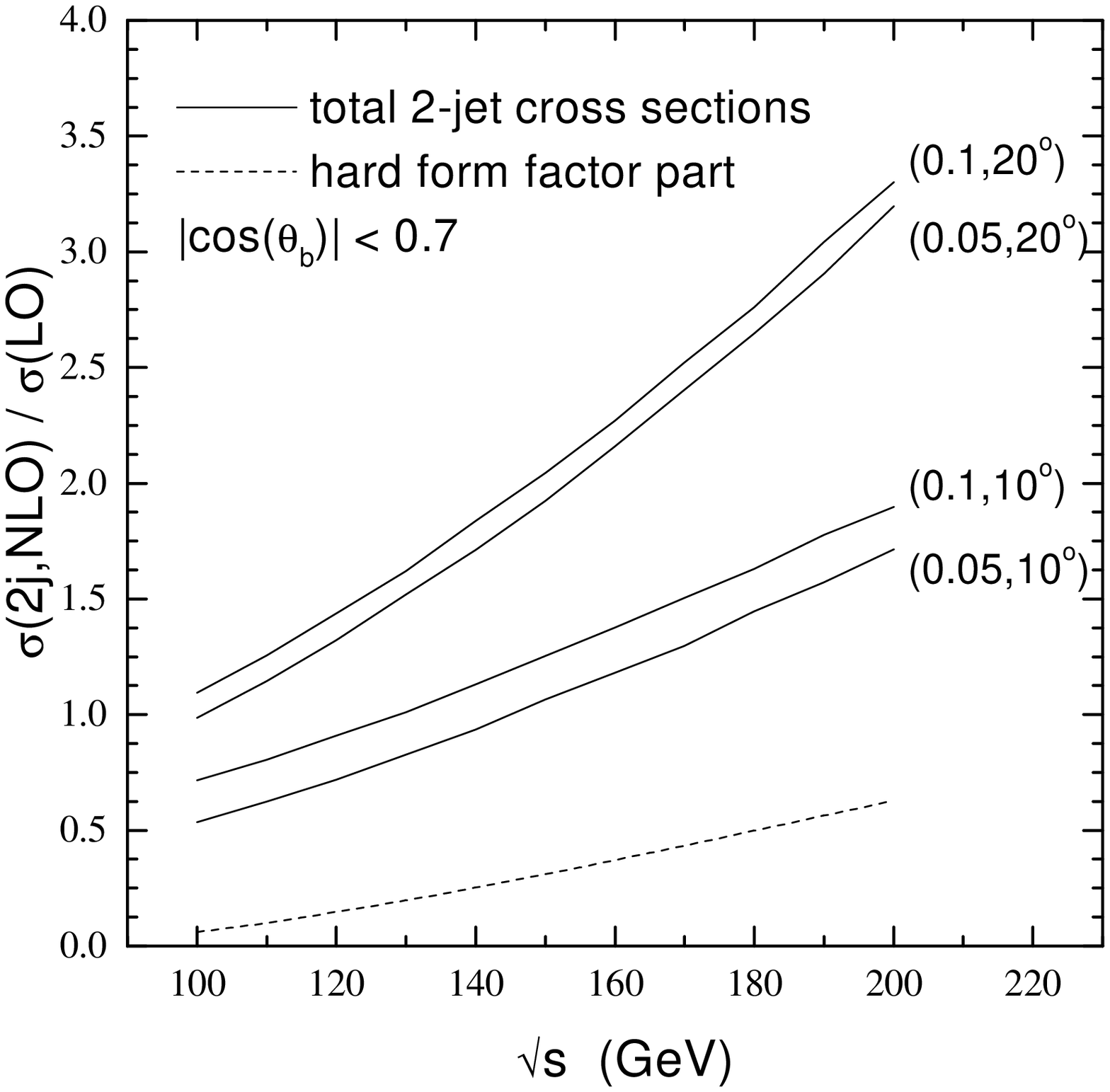,width=12cm}
\caption{The  total two--$b$--jet
cross section (i.e. exact next--to--leading order contribution
plus resummed hard form factor) normalized to leading order, 
for various jet parameters $(\epsilon,\delta)$.
Also shown (dashed line) is the hard form factor part alone.}
\label{fig:total}
\end{figure}

Our results are summarized in Fig.~\ref{fig:total}, where we display 
(solid curves) the total
two--jet cross section (i.e. exact next--to--leading order contribution
plus resummed hard form factor) for four choices of the parameters
$(\epsilon,\delta)$. Also shown (dashed curve) is the hard form factor
part alone. This is of course independent of the jet definition.
The form factor evidently gives a significant contribution to the total
cross section, especially for narrow jets.
From Fig.~1, we see also that using only the one--loop ($6{\cal F}$)
part of the hard form factor would result in a {\it negative} two--jet
cross section for narrow jets, a result first pointed out in \cite{jt1}.
With the resummed form factor, positivity and stability is restored. 

An additional potentially important higher--order
 contribution for large energies comes from the square of the {\it imaginary} part
 of the one--loop box diagram. An explicit calculation gives \cite{jt1}
\begin{equation}
\frac{d \sigma_{\rm Im}} {d \sigma_{\rm Born}}
\left( \gamma + \gamma \longrightarrow
q + \overline{q}, J_z = 0 \right) \approx \frac{\alpha_s^2}{18 \pi^2} \frac{s}{m_b^2}
\cos^2 \theta (1 - \cos^2 \theta ) \leq \frac{\alpha_s^2}{72 \pi^2} \frac{s}{m_b^2}
\label{eq:im}
\end{equation}
Note that, unlike $\sigma_{\rm Born}$, this contribution has no $m_q^2/s$ 
suppression and therefore eventually dominates the cross section
at very high scattering energy. For the energies considered here, however,
it is negligible.

What can we say about the corrections not included in
Fig.~\ref{fig:total}? Of course the unknown exact NNLO corrections
could be important, especially for the hard collinear contributions
to the two--jet cross section. We note here that, as at NLO, there
is no $m_q^2/s$ suppression of such terms, but tightening the
two--jet requirement will tend to reduce their significance.
We might again expect that the hard form factor is the dominant
part of the virtual/soft multigluon contributions. Note also
that as long as the parameter $\epsilon$ is not taken to be too small,
in which case large logarithms 
($\sim \alpha_s \log (m_q^2/s) \log  \epsilon$)
would require resummation,
the Sudakov form factor (see Eq.~(\ref{eq:vps})) should not
play a major role.
Finally, it is possible that the sub-leading (e.g. $\alpha_s^n \log^{2n-1}
(m_q^2/s)$) logarithmic
contributions to the hard form factor are important (although
this is not suggested by our exact
next--to--leading order studies). The calculation beyond NLO
would seem to be an extremely difficult task however \cite{ms}.

In conclusion, we have demonstrated the numerical importance
of the `hard' (non-Sudakov) $\log^2(m_q^2/s)$ contributions to the $J_Z=0$
$\gamma\gamma \to b \overline{b}$ cross section at typical photon
collider energies relevant for intermediate--mass Higgs searches.
 Our study, although it
includes for the first time all the
currently available theoretical information, is of course far 
from complete.\footnote{A more detailed phenomenological 
analysis will be presented elsewhere \cite{kms}.}
We have not, for example, included  running $\alpha_s$,
hadronization or  detector effects. These would require
dedicated Monte Carlo studies, such as those performed
in Ref.~\cite{bkso} for example. In this context, it is worth
pointing out that including the new hard logarithm form factor
in such studies should be straightforward.

\vspace{0.5cm}
\noindent{\bf Acknowledgements}\\ 
We would like to thank V.A.~Khoze for valuable discussions.
This work was supported in part by the EU Fourth Framework Programme `Training and Mobility of 
Researchers', Network `Quantum Chromodynamics and the Deep Structure of Elementary Particles', 
contract FMRX-CT98-0194 (DG 12 - MIHT).  
 
\appendix

\end{document}